\documentclass[a4paper,12pt]{article}
\pdfoutput=1

\usepackage[hyphens]{url}
\usepackage{lineno,hyperref}
\usepackage{xcolor}
\hypersetup{
   colorlinks,
   linkcolor={red!50!black},
   citecolor={blue!50!black},
   urlcolor={blue!80!black}
}

\usepackage{authblk}
\usepackage[natbibapa]{apacite}

\usepackage{todonotes}

\usepackage{amsmath}
\usepackage{amssymb}
\usepackage{color}

\usepackage{placeins}

\usepackage{multicol}
\usepackage{enumitem}

\usepackage{bm}

\usepackage[margin=0.5in]{geometry}

\begin{document}
 
\title{Regions of Interest as nodes of dynamic functional brain networks}

\author[1]{Elisa Ryypp\"o}
\author[2,3]{Enrico Glerean}
\author[4]{Elvira Brattico}
\author[1]{Jari Saram\"aki}
\author[1,3,*]{Onerva Korhonen}

\affil[1]{Department of Computer Science, School of Science, Aalto University, Espoo, Finland}
\affil[2]{Turku PET Centre, University of Turku, Turku, Finland}
\affil[3]{Department of Neuroscience and Biomedical Engineering, School of Science, Aalto University, Espoo, Finland}
\affil[4]{Center for Music in the Brain, Department of Clinical Medicine, Aarhus University, \& The Royal Academy of Music Aarhus/Aalborg, Denmark}
\affil[*]{Corresponding author: Onerva Korhonen, Aalto University, Department of Computer Science, P.O. Box 15400, FI-00076 Aalto, Finland, email: onerva.korhonen@aalto.fi}

\date{}
\maketitle

\begin{abstract}
The properties of functional brain networks strongly depend on how their nodes are chosen.
Commonly, nodes are defined by Regions of Interest (ROIs), pre-determined groupings of fMRI measurement voxels.
Earlier, we have demonstrated that the functional homogeneity of ROIs, captured by their \emph{spatial consistency}, 
varies widely across ROIs in commonly-used brain atlases.
Here, we ask how  ROIs behave as nodes of dynamic brain networks.
To this end, we use two measures: \textit{spatiotemporal consistency} measures changes in spatial consistency  across time and \textit{network turnover} quantifies the changes in the local network structure around a ROI.
We find that spatial consistency varies non-uniformly in space and time, which is reflected in the variation of spatiotemporal consistency across ROIs.
Further, we see time-dependent changes in the network neighborhoods of the ROIs, reflected in high network turnover. Network turnover 
 is  nonuniformly distributed across ROIs: ROIs with  high spatiotemporal consistency have low network turnover.
Finally, we reveal that there is rich voxel-level correlation structure inside ROIs. Because the internal structure and the connectivity of ROIs vary in time, the common approach of using static node definitions may be surprisingly 
inaccurate. 
Therefore,  network neuroscience would greatly benefit from node definition strategies  tailored  for dynamical networks.
\end{abstract}

\section{Introduction} \label{intro}

In 1909, Korbinian Brodmann published the results of his seminal work: maps of brain areas with different cytoarchitectures. His results
were among the first to suggest that the brain does not process information as an undivided entity. Instead, cognitive tasks are distributed among specialized brain areas. Since
Brodmann's time, the neuroscientific community has reached concensus on the distributed nature of brain function (see \citet{wig2011concepts} for a review). 
Information processing in the brain is based on the balance between segregation and integration: there are clusters with strong internal connections and weak long-range connectivity between them \citep{friston1994functional, tononi1994measure, sporns2013network}. 

Because of the crucial role of connectivity in the brain function, it is natural to model the brain as a network. In a network model of the brain, the nodes represent brain areas and the links represent the anatomical or functional connections between the nodes \citep{wig2011concepts, sporns2013network, sporns2013human, bassett2017network}.
Network neuroscience has unveiled several important features of the structure and function of the human brain.
For reviews, see for example
\citet{bassett2017network, betzel2016multi, sporns2013human, wig2011concepts}.

Networks of the brain vary across people and in time. Structural and functional brain networks have been reported to differ between people, in particular between diseased subjects and
healthy controls, and to change across the lifespan \citep{papo2014functional, bassett2009, sporns2013network, chan2017resting}. Functional brain networks vary on shorter timescales too,
for example with different cognitive tasks \citep{bassett2011dynamic, honey2007network, braun2015dynamic, gottlich2017viewing}. 
However, the traditional tools of connectivity analysis cannot capture this time variation: there is still a lack of appropriate methods for understanding the dynamics of brain networks.

There are two questions of fundamental importance for functional brain networks: what do the nodes represent, and how are their links defined? The common approach is to use Regions of Interest (ROIs) as the nodes. ROIs are 
collections of fMRI measurement voxels defined on the basis of anatomy, connectivity profiles, or function
(for a review, see \citet{dereus2013}). The BOLD time series of each voxel follows the changes in the voxel's level of activity. To arrive at a time series that represents an entire ROI, its voxel time series are typically averaged.
Then, the weights of the links between ROIs are quantified with some similarity measure of their time
series, such as the commonly used Pearson correlation coefficient.

The ROI time series are typically taken as accurate representations of the dynamics of the voxels within the ROI. Consequently, a minimum requirement for a ROI to be reasonably defined is its \emph{functional homogeneity}:
each of the voxels should have similar dynamics. In our previous work \citep{korhonen2017consistency}, we have used the concept of \textit{spatial consistency} for quantifying this functional
homogeneity. We found that spatial consistency varies widely across ROIs in the commonly-used parcellations, indicating that the assumption of functional homogeneity does not
hold for all ROIs in functional brain networks.

There are two possible reasons for low spatial consistency. First, it is possible that it indicates technical problems in the investigated parcellations: although functionally homogeneous regions may exist
in the brain, the parcellations are not able to capture these regions. Second, spatial consistency may vary in time: averaging over periods of extremely low and moderately high consistency would yield 
low values of average consistency. In \citet{korhonen2017consistency}, we have speculated that the variation of spatial consistency between ROIs may not be just a technical issue that can be overcome by some sophisticated parcellation scheme.
Instead, it may carry cognitive meaning and be related to changes in the ROIs' activation, for example.

In the present work, we generalize the investigation of spatial consistency into dynamic brain networks. We explore how spatial consistency
varies in time, and ask how its variation relates to changes in the local network structure around ROIs. To this end, we use two measures: \textit{spatiotemporal consistency} quantifies temporal changes in spatial consistency, 
and \textit{network turnover} 
measures the amount of turnover in a node's  network neighborhood across time.
We use in-house data collected from 13 healthy subjects
during free music listening and resting-state data of 28 healthy subjects from the Autism Brain Imaging Data Exchange (ABIDE) initiative
\citep{di2014autism}. The in-house dataset is a subset of a larger dataset that has been earlier partially described in \citet{alluri2015musical, burunat2015action, alluri2017connectivity}.

With these data, we show that the ROIs exhibit varying levels of spatiotemporal consistency, which indicates that their spatial consistency indeed changes in time. Further, significant turnover takes place in the 
neighborhoods of 
many ROIs. Network turnover is high especially for ROIs with low spatiotemporal consistency. Looking at the constituent voxels of ROIs in detail, we see that ROIs often have rich internal correlation structure 
that varies in time. 
 
These results indicate that the topology of functional brain networks changes continuously on short time scales, which should be taken into account
in brain network studies. Further, the significant temporal variation of functional homogeneity may suggest that new, dynamical ways of defining nodes are required for creating an accurate
network model of the brain. Importantly, the variation of functional homogeneity should not be seen as a technical issue that should be eliminated with some parcellation approach, but a phenomenon that carries cognitive meaning and that should be taken into account in the analysis of dynamic functional connectivity.

\section{Results}

\subsection{Spatial consistency of ROIs varies across time}

\begin{figure}[]
  \begin{center}
      \includegraphics[width=1\linewidth]{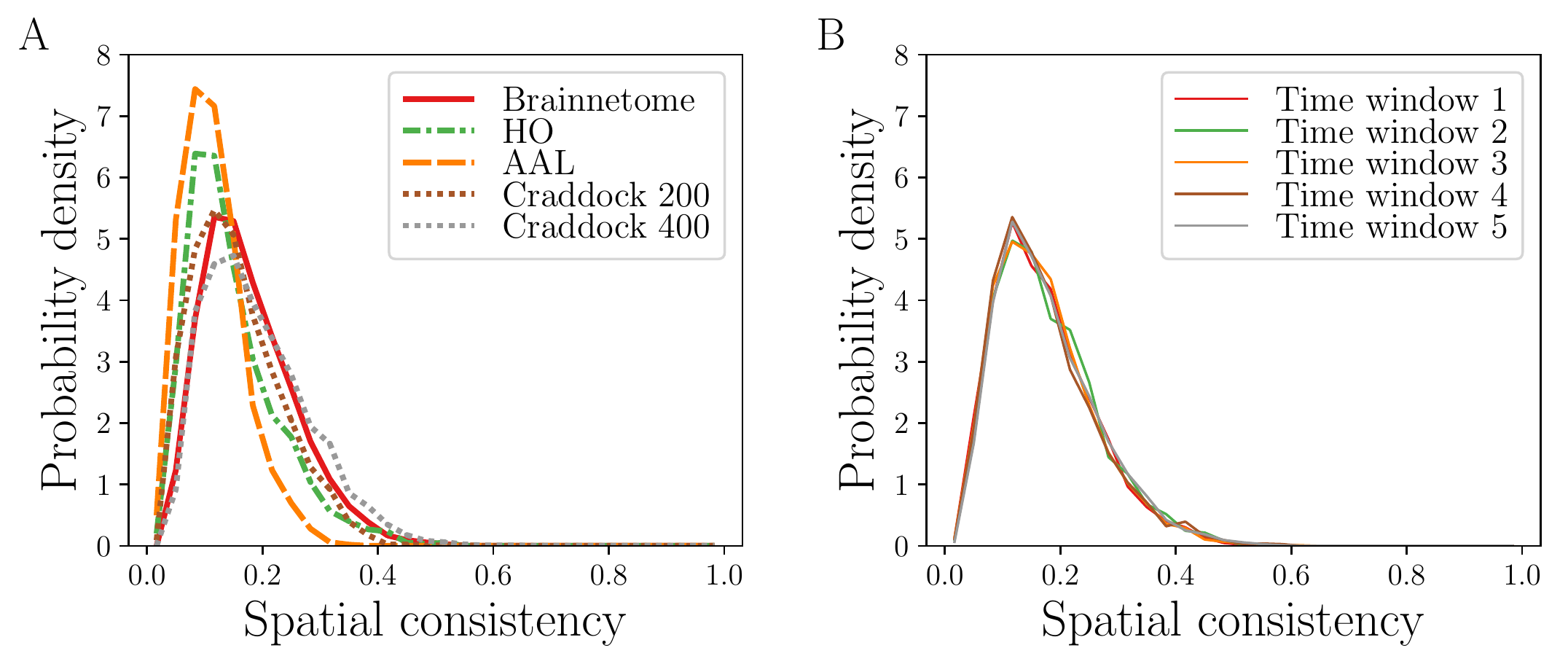} 
      \caption{The distribution of spatial consistency over ROIs indicates variation in functional homogeneity. A) Distributions of static spatial consistency for
      the five parcellations investigated. B) Distributions of spatial consistency calculated separately for five time windows of 80 samples for the Brainnetome ROIs. There is no visible
      difference between the distributions. All distributions have been calculated from the pooled
      data of 13 subjects. For AAL, HO, and Craddock 200/400, see Fig.~S1.}
      \label{fig:consistency-distributions-main}
 \end{center}
\end{figure}

Using pre-defined ROIs as nodes of functional brain networks is based on the assumption of functional homogeneity: all voxels within a ROI are assumed to have similar dynamics which can
be accurately represented by the ROI time series. To test this assumption, we calculated the distribution of \textit{spatial consistency} for five commonly-used parcellations of the brain: connectivity-based Brainnetome atlas and
Craddock 200/400 parcellations as well as two anatomical atlases: HarvardOxford (HO) and Automated Anatomical Labeling (AAL). 
Spatial consistency is defined as the average Pearson correlation
coefficient between the voxel time series in a ROI (see Eq.~\eqref{eq:spatial-consistency}). The results are in concordance with our earlier observations \citep{korhonen2017consistency}: although the maximum spatial 
consistency is moderately high 
(Brainnetome: $\phi_{spatial}=0.53$, HO: $\phi_{spatial}=0.53$, AAL: $\phi_{spatial}=0.34$, Craddock 200: $\phi_{spatial} = 0.55$, Craddock 400: $\phi_{spatial}= 0.65$), the distribution 
of spatial consistency is broad and peaks at low values (Brainnetome: $\phi_{spatial}=0.12$, HO: $\phi_{spatial}=0.083$, AAL: $\phi_{spatial}=0.083$, Craddock 200: $\phi_{spatial}= 0.12$, Craddock 400: $\phi_{spatial}=0.15$) 
(Fig.~\ref{fig:consistency-distributions-main}A). For Brainnetome or Craddock 200/400, there is no significant correlation between ROI size in voxels
and spatial consistency (Brainnetome: Pearson correlation coefficient $r=0.10$, $p=0.12$, Fig.~\ref{fig:results-roi-size}A; Craddock 200: $r=-8.31\times10^{-4}$, $p=0.991$; Craddock 400: $r=0.031$, $p=0.538$). For AAL and HO, there is a weak but significant correlation between ROI size and spatial consistency 
(AAL: $r=-0.32$, $p=4.13\times10^{-4}$; HO: $r=-0.33$, $p=8.62\times10^{-5}$).
The spatial consistency investigated here was calculated over the whole measurement time series; we will from here on refer to it as \textit{static spatial consistency}.

At least two different scenarios can explain the low values of static spatial consistency. On one hand, the voxels in a ROI may just have uncorrelated dynamics across the whole measurement time series. On the other hand, a moderately low level of correlation 
between the voxel time series may result from changes in the overall pattern, \emph{e.g.}~there may be periods of highly correlated activity and periods of no correlations at all. In the latter scenario, one would obtain time-dependent 
changes in spatial consistency by dividing the
measurement time series into shorter time windows. Therefore, we divided the measurement time series into five sliding windows 
of 80 samples each, with 50\% overlap between consecutive windows, and investigated the spatial consistency separately for each time window. 

We found no visible difference between distributions of spatial consistency calculated in different time windows (Fig.~\ref{fig:consistency-distributions-main}B).
One could assume that decreasing the time window length and
the overlap between consecutive windows would cause the spatial consistency distributions to differ more between windows. However, decreasing the window length to 50 samples and the overlap to 25\% did not induce more variation
between time windows (data not shown). On the other hand, increasing the overlap to the largest possible value, window length - 1, did not affect the distributions of spatial consistency either (see Fig.~S29).

\begin{figure}[]
  \begin{center}
      \includegraphics[width=1\linewidth]{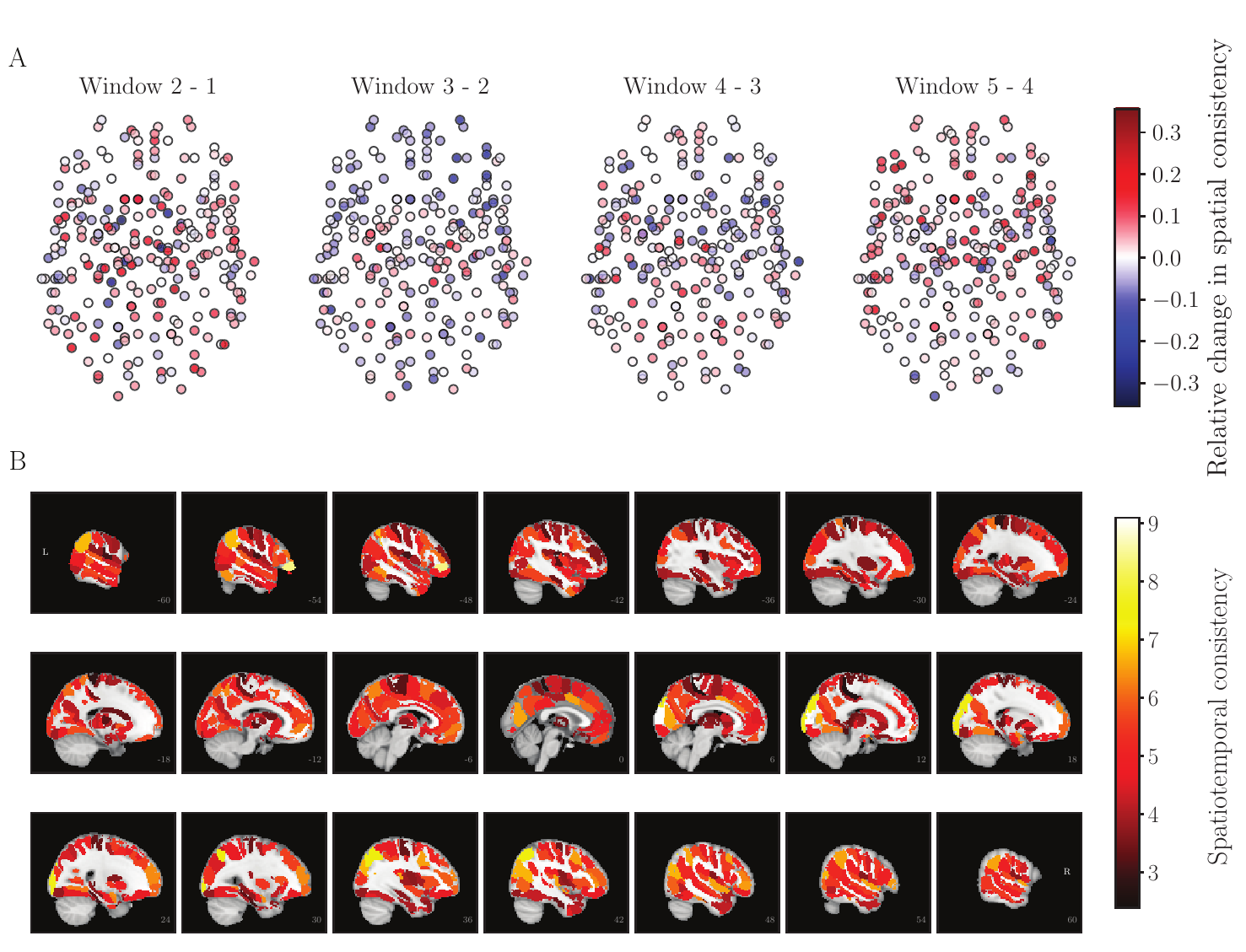} 
      \caption{Spatial consistency of ROIs varies between time windows. A) Relative changes of spatial consistency between consecutive time windows. Changes in spatial consistency are non-randomly distributed in time, meaning that changes between
    different time windows are not similar. Further, the changes show strong spatial correlations: the spatial consistency of anatomically adjacent ROIs tends to change in the same way. The location of nodes in the visualization is 
    determined by a two-dimensional projection of the anatomical coordinates of the Brainnetome ROI centroids. The visualization follows the neurological convention: the right hemisphere is on the right and the frontal areas in the upper part of the visualization. 
    B) Spatiotemporal consistency of the Brainnetome ROIs on the brain surface. As one may expect based on A), spatiotemporal consistency has a non-random anatomical distribution and shows strong spatial correlations. 
    All results are averages over 13 subjects. Grayscale areas are not included in the present study (white matter and cerebellum). For AAL, HO, and Craddock 200/400, see Figs.~S2, 
    S3, S4, and S5.}
    \label{consistency-results}
 \end{center}
\end{figure}

At the level of single ROIs, however, the situation is different. The
spatial consistency of most ROIs changes between time windows, and the largest relative changes in spatial consistency are around 30\% (Fig.~\ref{consistency-results}A). These changes have a non-random spatial distribution
and seem to occur in clusters larger than simple ROIs. This is visible for example as the drop in spatial consistency of the frontal regions between the time windows 2 and 3. A possible reason for the similar behavior of spatially
close ROIs is their assumed functional similarity; in particular, ROIs belonging to the same functional subsystem of the brain may be expected to behave similarly in terms of spatial consistency.

Similar time-dependent changes also take place 
in the spatial consistency ranks of ROIs (data not shown), demonstrating that the observations cannot be explained by changes in the overall level of spatial consistency. Note that in ROIs with small spatial consistency, even large
relative changes may be rather small in the absolute sense; therefore the distributions of spatial consistency obtained in different time windows are almost identical at lower consistency values (see Fig.~\ref{fig:consistency-distributions-main}B),
although we see relative changes in consistency of almost every ROI. At higher consistency values, large relative changes are larger  in the absolute sense as well. Therefore, distributions obtained in different time windows differ more from each other.

In order to quantify the amount of temporal variation in spatial consistency, we defined the spatiotemporal consistency as the inverse of the averaged relative change of spatial consistency across
time windows (see Eq.~\eqref{eq:stability-of-spatial}). In other words, spatiotemporal consistency is a static measure that tells how stable the ROI's spatial consistency is over time windows on average.

Spatiotemporal consistency is not anatomically uniform (Fig.~\ref{consistency-results}B), and identity of the ROIs with the highest and lowest spatiotemporal consistency varies largely across subjects. This inter-subject difference
is partly explained by technical issues. The investigated parcellations have been defined at the group level, and they match differently with the individual anatomy and brain function of different subjects. On the other hand, differences
in spatiotemporal consistency may also reflect the different cognitive responses of different subjects during free music listening.

The Brainnetome ROIs with the highest average spatiotemporal consistency include the right cuneus (5\_3), left inferior frontal gyrus (6\_4), right occipital gyrus (4\_3), right superior occipital gyrus (2\_1) and right inferior parietal lobule (6\_2).
In AAL, among the ROIs with the highest average spatiotemporal consistency are the
left medial orbitofrontal cortex, right cerebellar area 6, left middle temporal gyrus, right insula and left gyrus rectus.
In HO, the ROIs with the highest average spatiotemporal consistency include the right supracalcarine cortex, left frontal pole, anterior division of left superior temporal gyrus, right angular gyrus and posterior division of right middle temporal gyrus.
In Craddock 200/400, ROI boundaries do not respect anatomical landmarks, and the ROIs are referred to by only numbers. For the location of the Craddock 200/400 ROIs with the highest and lowest average spatiotemporal
consistency, the reader is referred to Figs.~S18 and S19.

In Brainnetome,  the ROIs with the lowest average spatiotemporal consistency are the right parahippocampal gyrus (6\_5 and 6\_2), right paracentral lobule (2\_1), left postcentral gyrus (4\_4 and 4\_2).
In AAL, the ROIs with the lowest average spatiotemporal consistency include the left cerebellar area 4\_5, right supplementary motor area, left paracentral lobule, right parahippocampal gyrus and right thalamus.
In HO, the ROIs with the lowest averege spatiotemporal consistency include the right hippocampus, posterior division of right parahippocampal gyrus, right precentral gyrus, right thalamus and brain stem.
In all investigated atlases, many subcortical areas are among the ROIs with low spatiotemporal consistency. We will discuss possible reasons for this later in this article.  

As one possible explanation for the variation of spatiotemporal consistency across ROIs,
we found a weak but significant correlation between the ROI size and the spatiotemporal consistency in the Brainnetome atlas ($r=0.24$, $p=1.37\times10^{-4}$; see Fig~\ref{fig:results-roi-size}B). In AAL, HO, or Craddock atlases, there is no significant correlation between the spatiotemporal consistency and ROI size 
(AAL: $r=0.16$, $p=0.0963$; HO: $r=-0.025$, $p=0.770$; Craddock 200: $r=0.028$, $p=0.698$, Craddock 400: $r=0.014$, $p=0.780$). The lack of correlation in the Craddock parcellations is not surprising: these parcellations
aim at minimizing the variation of ROI sizes and they have clearly smaller SDs of ROI size than the other parcellations investigated. There are a few possible reasons for why we observe a correlation in the Brainnettome but not in AAL or HO. First, Brainnetome contains more ROIs than AAL or HO. Second, unlike Brainnetome, AAL and HO contain the cerebellum where ROIs are small due to anatomical reasons but they do not have systematically higher values 
of spatiotemporal consistency than the ROIs of the cerebral cortex. Finally, in AAL and HO the ROI size is negatively correlated with static spatial consistency; this is not the case for Brainnetome. On the other hand, in all
 atlases, ROIs with high static spatial consistency tend to have high spatiotemporal consistency as well (see below). Because of this, one would expect to see a \textit{negative} correlation between ROI size and spatiotemporal
consistency in HO and AAL. This negative correlation may have masked the positive correlation obtained for the Brainnetome atlas.

A sliding window with a 1 TR shift is commonly used for studying dynamic functional connectivity \citep{keilholz2017time}. In this approach, the overlap between consecutive time windows is as large as possible: window length - 1. 
In our case, however, this large an overlap would hide the changes in local network structure. It would also lead to extremely low values of network turnover (see below).
However, we investigated how the 1 TR shift  would affect the observed values of spatiotemporal consistency in the Brainnetome parcellation. As expected, using the 1 TR shift sliding window moved the distribution of spatial consistency slightly to the right
(distribution peaking at $\phi_{st} = 3.1$ vs $\phi_{st} = 4.9$; see Fig.~S29B): as the overlap between consecutive time series increases, there is less room for changes in spatial consistency. However, the 1 TR shift approach did not affect the overall shape
of the distribution of spatiotemporal consistency, and low values of spatiotemporal consistency that indicate large relative changes in spatial consistency are observed with this approach too. 

Subject motion is known to possibly affect the structure of functional brain networks \citep{power2012spurious}. Therefore,
one may ask if the temporal variation in spatial consistency is of genuine neurophysiological origin or if it could be explained by motion artifacts. To answer this, we investigated the temporal correlation between the mean
framewise displacement (FD) and the spatial consistency concatenated across subjects. However, they did not correlate significantly for any ROI in any of the investigated atlases. The correlation between
the static spatial consistency and the mean FD over subjects was not significant neither.

Temporal fluctuations in functional connectivity of the brain may underlie changes in cognitive processing \citep{cocchi2017neural}. We found a significant correlation between time-resolved functional connectivity \citep{zalesky2014time, cocchi2017neural}
and spatial consistency for some ROIs of Brainnetome and HO. For further details, the reader is referred to Supplementary Results.

Throughout the
present article, we investigated five atlases: Brainnetome, Craddock 200/400, AAL, and HO. Despite the differences between these atlases, we obtained highly similar results for all of them. In the main article, we concentrate on the
results obtained with the Brainnetome atlas; for detailed results and visualizations for the Craddock 200/400, AAL, and HO atlases, the reader is referred to Supplementary Results.

To verify that the results generalize, we repeated all analyses for a second, independent dataset from the ABIDE I initiative
\citep{di2014autism}. The results obtained using the ABIDE data were very similar to those reported here; full details can be found in Supplementary Results.

\subsection{Network neighborhoods of nodes change in time}

\begin{figure}[]
  \begin{center}
      \includegraphics[width=1\linewidth]{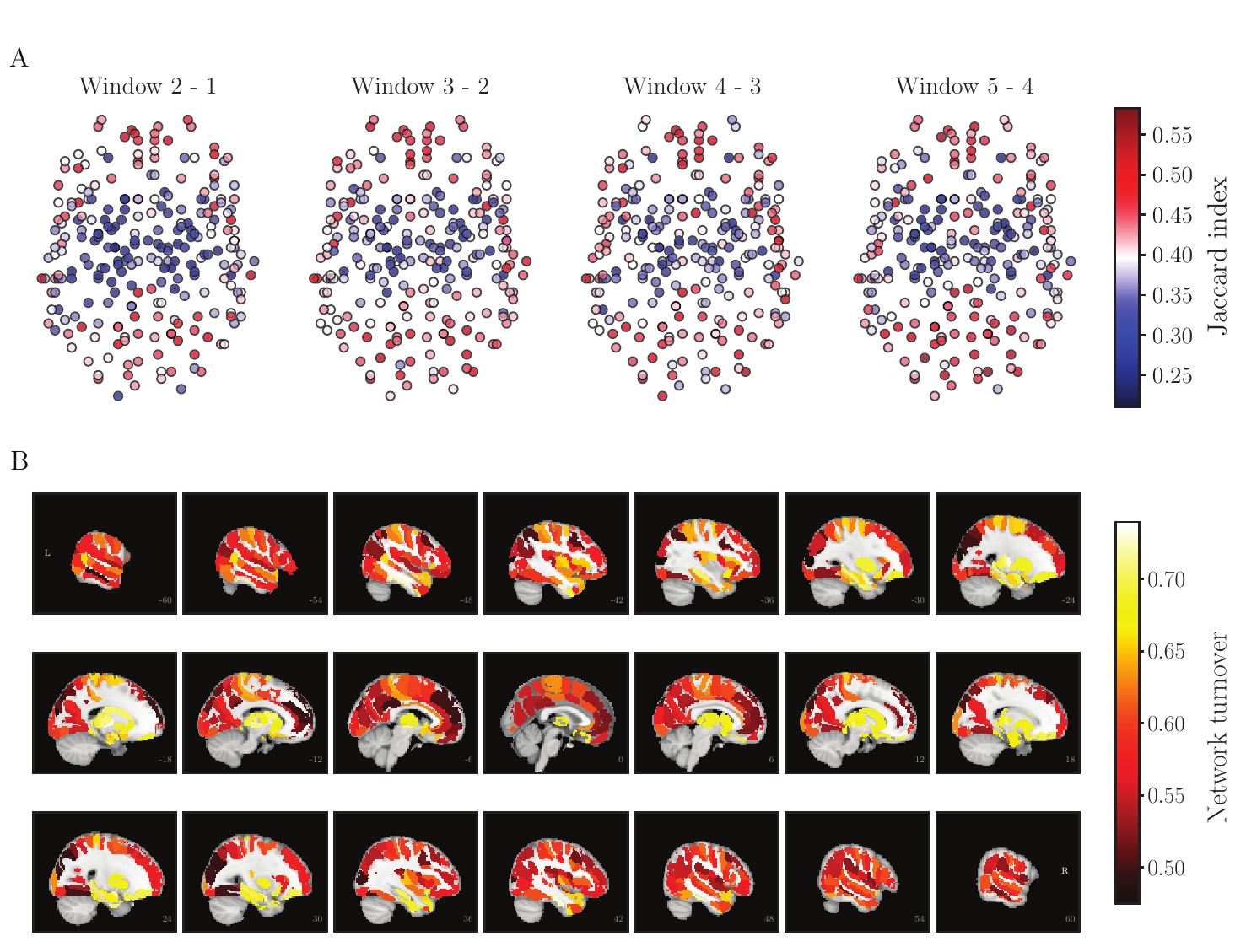} 
    \caption{There is strong neighborhood turnover in dynamic functional brain networks. A) The Jaccard index between consecutive time windows. Values of the Jaccard index are nonuniformly distributed in space and show strong spatial 
    correlations: adjacent ROIs show similar values. Node locations are as in Fig.~\ref{consistency-results}. B) Network turnover on the brain surface in the Brainnetome atlas. High network turnover of 
    subcortical ROIs as compared to cortical ROIs is particularly visible. Jaccard index values and network turnovers have been averaged over 13 subjects. Fo AAL, HO, and Craddock 200/400, see 
    Figs.~S6, S7, S8, and S9.}
    \label{turnover-results}
 \end{center}
\end{figure}

The structure of functional brain networks is known to change in time. For an individual node, this means that the local structure around the node, \textit{i.e.} the identity of its neighbors, may change.
This change can be quantified in terms of the Jaccard index 
between the node's sets of neighbors in consecutive time windows. We defined a ROI's closest neighborhood 
as its 35 most strongly linked 
neighbors and investigated the Jaccard index. Indeed, we found significant changes in ROIs' neighborhoods in time (Fig.~\ref{turnover-results}A). ROIs with the highest neighborhood turnover may
change up to 75\%  of their closest neighbors between two time windows, corresponding to a Jaccard index of 0.25. Even the ROIs with the most stable neighborhoods reach a Jaccard index value of 0.55 only, meaning that half of their closest
neighborhood changes between consecutive time windows.  For comparison,
 shuffling the weights of 5\% of randomly chosen links in the full network for 1000 times yields an average Jaccard index of 0.89$\pm$0.062 (mean$\pm$SD).

The spatial distribution of the Jaccard index over the ROIs appears stable in time. In particular, subcortical ROIs have clearly lower
Jaccard index values than cortical ROIs independent of the time window pair investigated. We will discuss possible reasons for this later.

\textit{Network turnover}, defined as the complement of the mean Jaccard index of the ROI's neighborhood over time (see Eq.~\eqref{eq:network-consistency}), 
quantifies the overall tendency of the neighborhood to change in time. Like spatiotemporal consistency, network turnover
varies between ROIs (Fig.~\ref{turnover-results}B). As expected due to their low Jaccard index values, subcortical ROIs have higher network turnover than cortical ROIs. Network turnover is spatially strongly correlated: anatomically 
adjacent ROIs tend to have similar values of network turnover.

The Brainnetome ROIs that have the highest average network turnover include the left and right parahippocampal gyrus (6\_5), left thalamus (8\_2), and right and left inferior temporal gyrus (7\_1).
In AAL, these include vermis 9, the left caudate nucleus, left cerebellar area 3, vermis 1\_2, and right olfactory cortex.
In HO, the ROIs with the highest average network turnover include the right and left pallidum, anterior division of left temporal fusiform cortex, vermis X,  and vermis VIIIb. For the Craddock 200/400 ROIs with the highest and
lowest network turnover, the reader is referred to Figs.~S18 and S19.
There is some variation in the identity of the 
highest network turnover ROIs across subjects; however, subcortical areas tend to have high network turnover in all subjects.

Identity of ROIs with the lowest network turnover vary a lot across subjects.
The ROIs with the lowest average network turnover include in Brainnetome the left occipital gyrus (4\_1), left middle temporal gyrus (4\_1), right superior occipital gyrus (2\_2), and 
left superior frontal gyrus (7\_7 and 7\_3). In AAL, they include the right fusiform cortex, right cerebellar area 6, right superior occipital gyrus, left angular gyrus, and right middle occipital gyrus.
In HO, the low average network turnover ROIs include the left frontal pole, left middle frontal gyrus, left angular gyrus, left paracingulate gyrus, and left cuneal cortex.

\begin{figure}[]
  \begin{center}
      \includegraphics[width=0.5\linewidth]{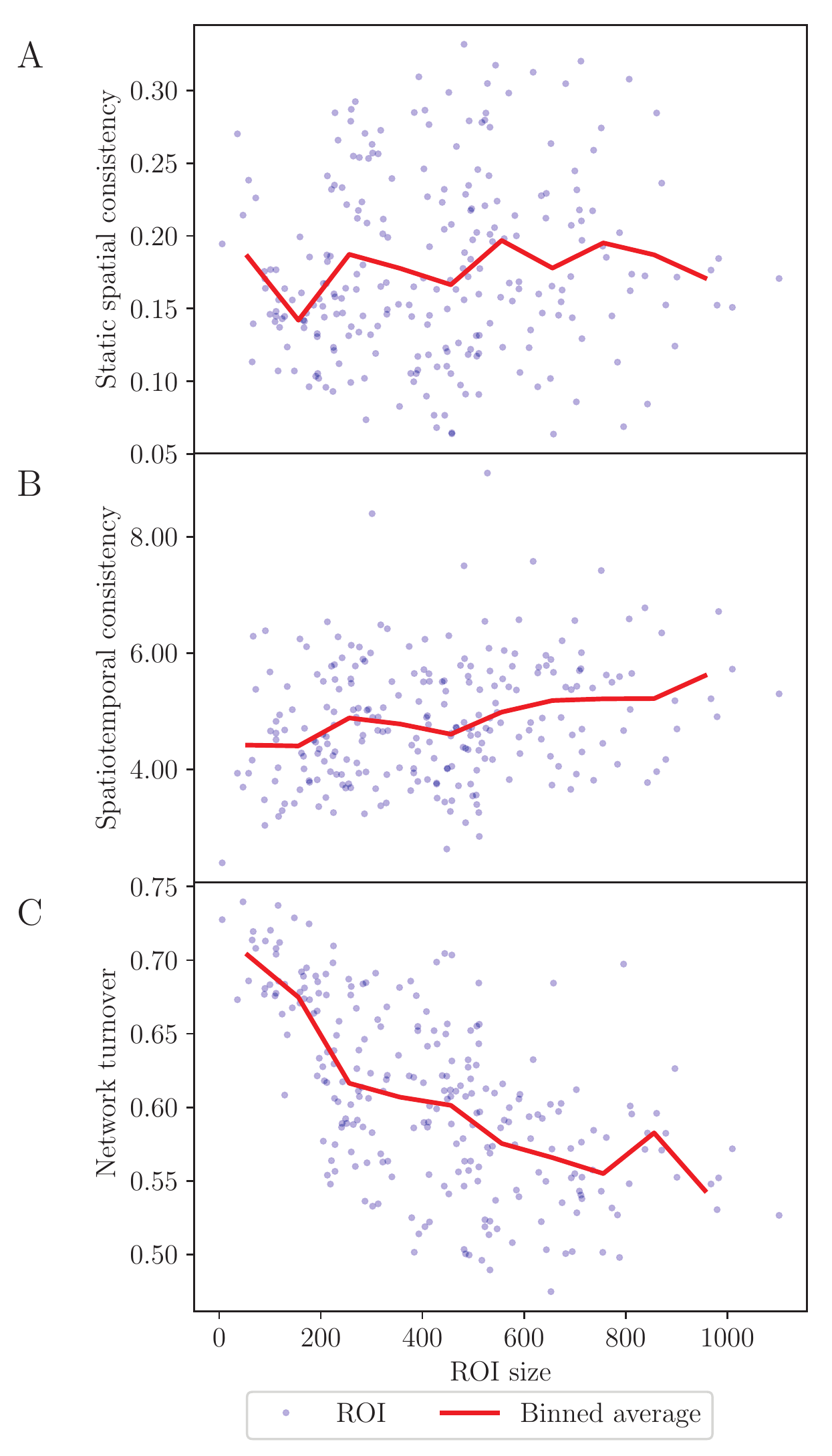} 
      \caption{Relationship of the sizes of the Brainnetome ROIs to their spatial and spatiotemporal consistency and network turnover. A) Static spatial consistency does not correlate with ROI size. B) There is a weak positive 
      correlation between
      spatiotemporal consistency and ROI size. C) Network turnover and ROI size are clearly negatively correlated. Data have been averaged over 13 subjects. The solid red lines show
      bin averages; binning has been done on the basis of ROI size. For AAL, HO, and Craddock 200/400, see Figs.~S10 and S11.}
      \label{fig:results-roi-size}
 \end{center}
\end{figure}

In addition to spatial variation, we found significant negative correlation
between ROI's size and network turnover (Brainnetome: $r=-0.60$, $p\ll10^{-5}$; HO: $r=-0.30$, $p=4.02\times10^{-4}$; AAL: $r=-0.42$, $p\ll10^{-5}$; Craddock 200: $r=-0.21$, $p=0.00232$; Craddock 400: $r=-0.41$, $p\ll10^{-5}$; Fig.~\ref{fig:results-roi-size}C). This correlation is most probably dominated
by the very high network turnover values of the subcortical ROIs that, for anatomical reasons, tend to be smaller than cortical ROIs. In the AAL and HO atlases, the correlation may have been partly shadowed by the lower number of ROIs
and the presence of cerebellar ROIs that are small but do not have systematically lower network turnover values than ROIs of the cerebral cortex.

\subsection{The ROIs with the highest spatiotemporal consistency have the lowest turnover in their neighborhood}\label{results:spatiotemporal-vs-network-consistency}

\begin{figure}[]
  \begin{center}
      \includegraphics[width=0.5\linewidth]{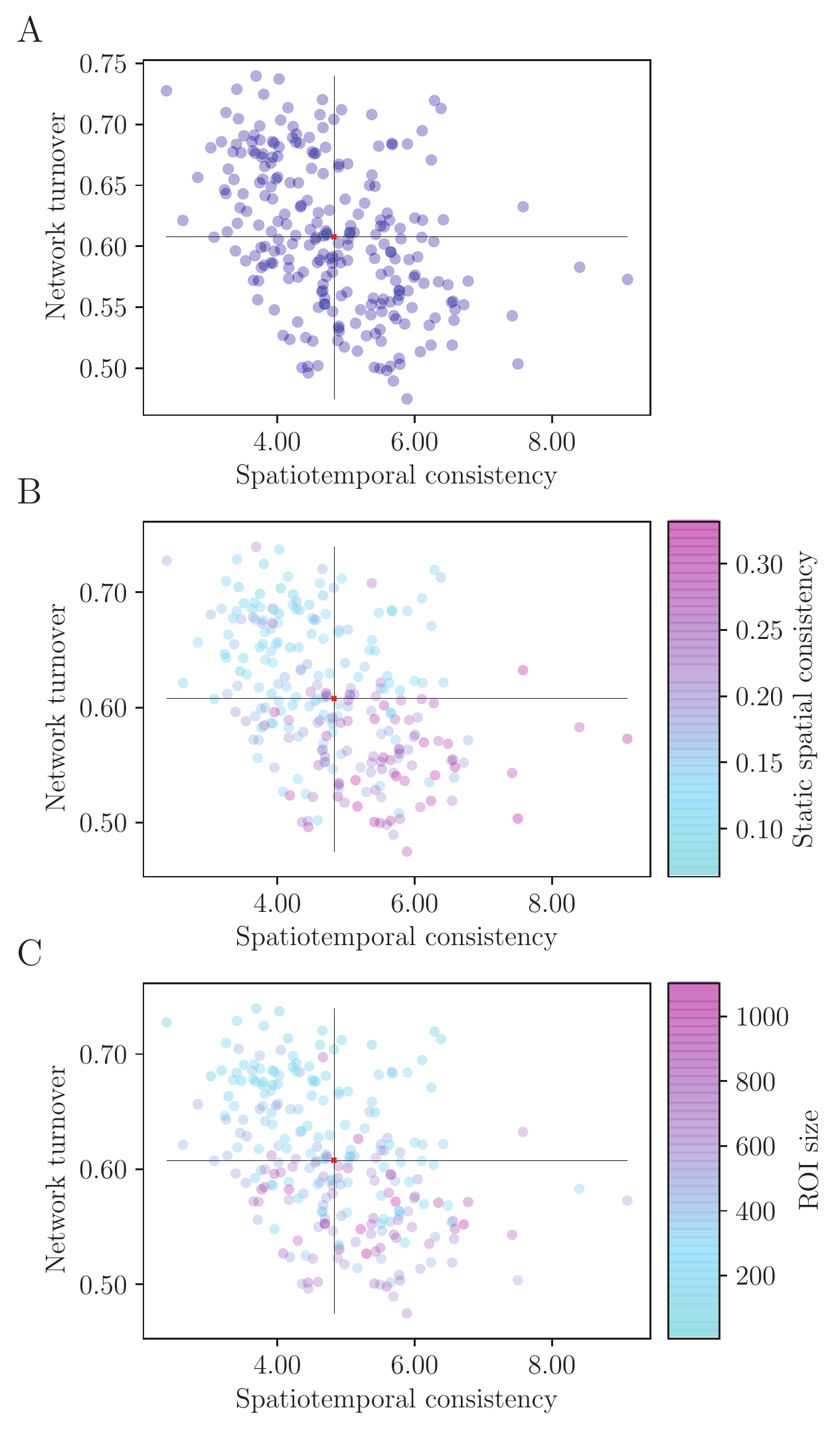} 
      \caption{Spatial and spatiotemporal consistency and network turnover depend on each other. A) Spatiotemporal consistency is negatively correlated with network turnover. B) ROIs with the highest
     static spatial consistency also have the highest spatiotemporal consistency and lowest network turnover. C) Largest ROIs tend to have highest spatial and spatiotemporal consistency and lowest network 
     turnover in the Brainnetome atlas. Data have been averaged over 13 subjects. For AAL, HO, and Craddock 200/400, see Figs.~S12, S13,
     S14, and S15.}
      \label{fig:results-scatter}
 \end{center}
\end{figure}

Next, we asked how spatiotemporal consistency and network turnover relate to each other. At the group level, \textit{i.e.} averaging the spatiotemporal consistencies and turnovers over subjects, we found a significant negative correlation between these measures 
(Brainnetome: $r=-0.42$, $p\ll10^{-5}$; HO: $r=-0.44$, $p\ll10^{-5}$; AAL: $r=-0.38$, $p=2.08\times10^{-5}$; Craddock 200: $r=-0.46$, $p\ll10^{-5}$; Craddock 400: $r=-0.42$, $p\ll10^{-5}$)
(Fig.~\ref{fig:results-scatter}A). In other words, ROIs with the highest spatiotemporal consistency have the lowest amount of turnover in their neighborhoods. These ROIs also have the highest static spatial consistency
(Fig.~\ref{fig:results-scatter}B).

The correlation between spatiotemporal consistency and network turnover was also visible, albeit weaker, at the level of single subjects (Brainnetome: significant ($p<0.05$) negative correlation for 6 subjects, negative but non-significant correlation
for 5 subjects, non-significant positive correlation for 1 subject; HO: significant negative correlation for 7 subjects, negative but non-significant correlation for 5 subjects, non-significant positive correlation for 1 subject;
AAL: significant negative correlation for 4 subjects, negative but non-significant correlation for 5 subjects, non-significant positive correlation for 4 subjects). A plausible reason for the weaker and less significant correlations
obtained for AAL is the number of data points: AAL contains fewer ROIs (116) than Brainnetome (246) or HO (138), which may have made it more difficult to obtain a significant correlation.

The connectivity profiles of ROIs with low network turnover change only little between time windows and they resemble the connectivity profiles obtained over the whole time series. Therefore, ROIs with low network turnover should have
stronger links in the networks extracted from the whole time series. As low-turnover ROIs tend to have high static spatial consistency, it is not too surprising that we found a correlation between static spatial consistency 
and degree in networks extracted from the whole time series. At 2.5\% density, this correlation was significant for all investigated parcellations (Brainnetome: $r=0.48$, $p\ll10^{-5}$; AAL: $r=0.31$, $p\ll10^{-5}$; HO: $r=0.41$,$p\ll10^{-5}$; Craddock 200: $r=0.50$, $p\ll10^{-5}$; Craddock 400: $r=0.57$, $p\ll10^{-5}$). The correlation remained significant also for higher network densities; the highest density where the correlation was present varied between parcellations 
(Brainnetome: $d=45.0$, $r=0.037$, $p=0.0368$; AAL: $d=10.0$, $r=0.11$, $p\ll10^{-5}$; HO: $d=10.0$, $r=0.057$, $p=0.0155$; Craddock 200: $d=30.0$, $r=0.058$, $p=0.00309$; Craddock 400: $d=40.0$, $r=0.0603$, $p=1.67\times10^{-5}$).

In Brainnetome, the ROIs with the highest spatiotemporal consistency and lowest network turnover tend to be
larger than ROIs with lower spatiotemporal consistency and higher network turnover (Fig.~\ref{fig:results-scatter}C). This is as one may expect based on the correlations between spatiotemporal consistency and ROI size, and 
network turnover and ROI size (Fig.~\ref{fig:results-roi-size}B,C). In AAL, HO, and Craddock 200/400, this relationship is less clear (see Figs.~S12C, S13C).

The relationship between spatiotemporal consistency and network turnover strongly depends on how we define spatiotemporal consistency. The definition given in Eq. \eqref{eq:stability-of-spatial} measures relative changes in 
spatial consistency. To get a more complete picture, we investigated also an alternative definition of spatiotemporal consistency that measures absolute changes. For details, the reader is referred to Supplementary Results.

\subsection{ROIs can be divided into two extreme groups on the basis of consistency and turnover, and these match with cortical and subcortical regions}\label{results:extreme-groups}

\begin{figure}[]
  \begin{center}
      \includegraphics[width=1\linewidth]{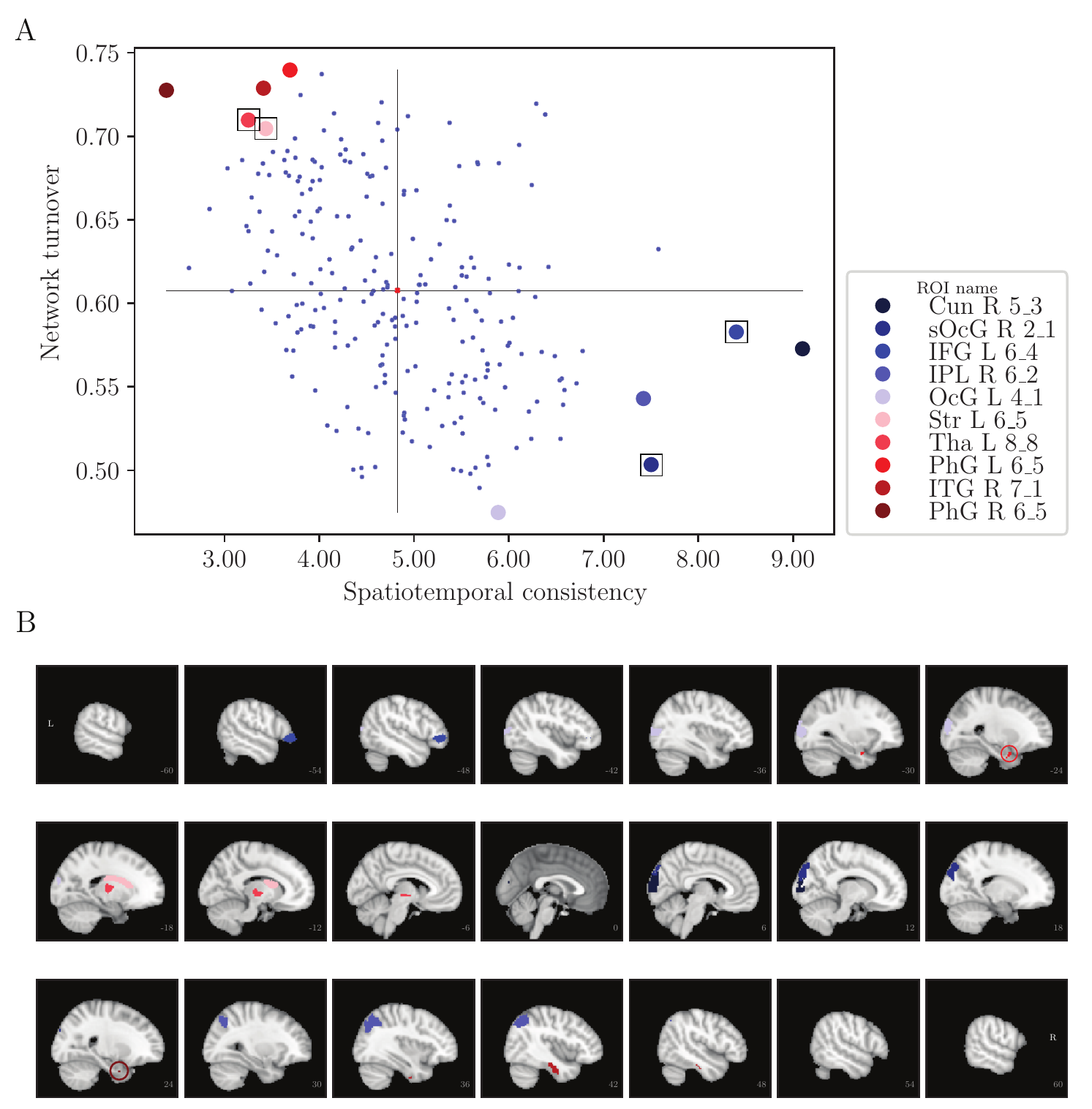} 
    \caption{Extreme ROIs in terms of spatiotemporal consistency and network turnover. A) Location of extreme ROIs in the space spanned by spatiotemporal consistency and network turnover. The blue and red groups have been chosen 
    with the help of PCA (see text). The ROIs in the blue group have high spatiotemporal consistency 
    and low network turnover, whereas the ROIs in the red group have low spatiotemporal
    consistency and high network turnover. The internal voxel-level connectivity of ROIs marked with a square is investigated in detail, see Fig.~\ref{fig:results-corrmats}. B) Location of extreme ROIs on the brain surface. 
    L: left, R: right, Cun: cuneus, sOcG: superior occipital gyrus, IFG: inferior frontal gyrus, IPL: inferior parietal lobule, OcG: occipital gurys, Str: striatum, Tha: thalamus, PhG: parahippocampal gyrus, ITG: inferior temporal
    gyrus. For AAL, HO, and Craddock 200/400, see Figs.~S16, S17, S18, and S19.}
    \label{top-roi-results}
 \end{center}
\end{figure}

So far, we have investigated the relationship between spatiotemporal consistency and network turnover at the population level. Next, we asked which specific ROIs are the ones with the highest and lowest values of spatiotemporal
consistency and network turnover. To this end, we obtained two groups of extreme ROIs by applying principal component analysis (PCA) in the space spanned by spatiotemporal consistency and network turnover. The extreme groups
contain the five ROIs with the largest and smallest projected coordinates on the first principal component. ROIs of the first group have high spatiotemporal consistency and low network turnover, and ROIs of the second 
group have lower spatiotemporal consistency and high network turnover (Fig.~\ref{top-roi-results}). As the PCA has only two degrees of freedom, the extreme groups could in principle have been defined by visual inspection alone; the
main reason for applying PCA was to avoid subjectivity and to ensure that the extreme groups are defined similarly in all investigated parcellations.

In Brainnetome, the first  group contains the right cuneus (5\_3), right superior occipital gyrus (2\_1), left inferior frontal gyrus (6\_4), right inferior parietal lobule (6\_2) and left occipital gyrus (4\_1).
The AAL ROIs of this  group are the right cerebellar area 6, left medial orbitofrontal cortex, right superior occipital gurys, left angular gyrus and right middle occipital gyrus.
In HO, this group comprises the left frontal pole, right and left supracalcarine cortex,  left middle frontal gyrus and right angular gyrus.
For Craddock 200/400 ROIs belonging to the extreme groups, the reader is referred to Figs.~S18 and S19.

The second group, \textit{i.e.} the ROIs with low spatiotemporal consistency and high network turnover, contains in Brainnetome the left and right parahippocampal gyrus (6\_5), right inferior temporal gyrus (7\_1), 
left thalamus (8\_8) and left striatum (6\_5).
The AAL ROIs that belong to this group are the right globus pallidum, left paracentral lobule, right olfactory cortex, right cerebellar area 9 and Vermis 1\_2.
In HO, this group contains the right and left pallidum, brain stem, right hippocampus and right thalamus.
In all five parcellations, most ROIs of this group are relatively small areas located deep in the brain. Because of the location, the signal-to-noise ratio (SNR) of the fMRI measurement tends to be low for these
areas. This may at least partially explain their low spatiotemporal consistency and may also limit the accuracy of estimating their network connectivity, leading to noisy closest neighborhoods and high turnover.

The sets of extreme ROIs in different parcellations are not the same, but this is to be expected. First, the ROIs of different parcellations have different shapes, sizes, and locations. Second, there are many ROIs with 
spatiotemporal consistency and network turnover that are rather close to those of the five most extreme ROIs; this hard threshold is of course arbitrary.

\subsection{Nontrivial, dynamic voxel-level structure occurs within ROIs}\label{results:internal-connectivity}

\begin{figure}[h!]
  \begin{center}
      \includegraphics[width=1\linewidth]{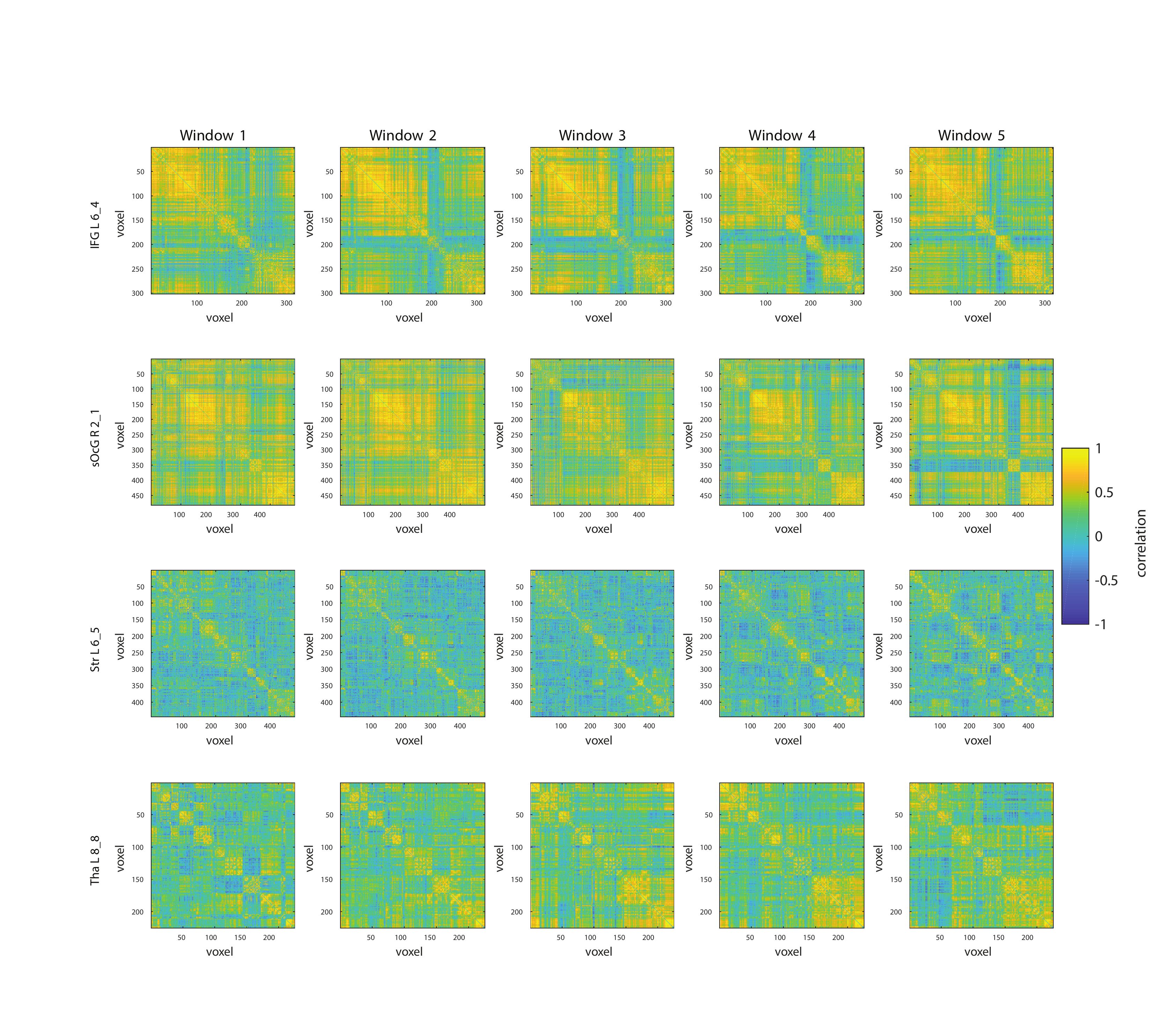} 
      \caption{The internal connectivity structure of ROIs is visible in the voxel-level correlation matrices of their internal connections. This internal structure changes in time. The upper two rows display matrices for 
      high-spatiotemporal-consistency-low-network-turnover ROIs, and the two lower rows those for low-spatiotemporal-consistency-high-network-turnover ROIs.  
      To order the voxels within each ROI, voxels were assigned to communities with the generalized Louvain method for multiplex networks, and then the Hamming distance between 
      these community assignments was used to find 
      the optimal leaf order of the hierarchical clustering tree  \citep{jeub2011generalized, mucha2010community}.
      The order of voxels is same in all time windows. Data of one representative subject are shown here. 
      L: left, R: right, IFG: inferior frontal gyrus, sOcG: superior occipital gyrus, Str: striatum, Tha: thalamus. For AAL, HO, and Craddock 200/400, see Figs.~S20 and S21.}
      \label{fig:results-corrmats}
 \end{center}
\end{figure}

From both groups of extreme ROIs, we selected two ROIs for a more detailed investigation. We chose the most extreme ROIs that were not exceptionally small or too large for the visualization discussed below. In Brainnetome, 
the selected ROIs were
the left inferior frontal gyrus (6\_4) and right superior occipital gyrus (2\_1) from the high-spatiotemporal-consistency-low-network-turnover group and the left striatum (6\_5) and left thalamus (8\_8)
from the opposite group. We calculated voxel-level correlation matrices to reveal the internal correlation structure inside these ROIs (Fig.~\ref{fig:results-corrmats}). 

The two groups are visibly different in terms of their correlation matrices: the overall correlation level is clearly higher for ROIs with high spatiotemporal consistency than for ROIs with low spatiotemporal consistency. The voxel-level 
correlations are not, however, uniformly distributed. Instead, a division into several internally highly correlated subareas is visible inside ROIs with high spatiotemporal consistency and  ROIs with low spatiotemporal consistency. 

This internal structure of ROIs is seen to change in time. In the right superior occipital gyrus that has high spatial and spatiotemporal consistency, the voxels are uniformly correlated across the whole ROI in time windows
1 and 2 but separate into two clusters between time windows 2 and 3. Similarly, the left thalamus that has low spatial and spatiotemporal consistency
shows time-dependent internal cluster structure. 

The internal structure of ROIs may affect their spatiotemporal consistency and static spatial consistency in several ways. For example, stable internal structure should 
manifest itself as high spatiotemporal consistency, because the average voxel-level correlation does not change in time. Fewer and larger subareas lead to larger amounts 
of correlated voxels within the ROI and should therefore be associated with higher static spatial consistency. On the other hand, a large number of small subareas should lead to low static spatial consistency, as should
a total lack of internal structure.

\section{Discussion}

\subsection{Functional homogeneity of ROIs varies in time}

The use of ROIs as nodes of fMRI brain networks assumes functional homogeneity: each of the ROI's voxels is thought to have similar dynamics, and therefore the ROI time series is considered as an accurate representation of the voxel-level
dynamics. Earlier \citep{korhonen2017consistency}, we have shown that this assumption does not hold for the ROIs of commonly-used parcellations. To this end, we used \textit{spatial consistency},
a measure of functional homogeneity defined as the mean Pearson correlation coefficient between voxel time series inside a ROI.

Functional homogeneity is often considered as a static ROI property. However, functional brain networks change in time, even on short time scales \citep{bassett2011dynamic, honey2007network, gottlich2017viewing}. Here, we 
investigated the temporal behavior of spatial consistency.  We divided fMRI data measured during a free music listening task into time windows and calculated
the relative change in spatial consistency between them. 
For quantifying the temporal variation in spatial consistency, we introduced  \textit{spatiotemporal consistency} as the inverse of the mean relative 
change in spatial consistency over time windows (see Equation~\eqref{eq:stability-of-spatial}).
We found that spatial consistency changes significantly in time, 
the largest relative changes being up to 30\%, resulting in low spatiotemporal consistency.

The concept of dynamic functional connectivity has been recently debated among the neuroscientific community. While many studies have reported time-dependent changes in the structure of functional brain networks
\citep{zalesky2014time, cocchi2017neural, bassett2011dynamic, honey2007network}, the neurophysiological meaning of these changes is not fully understood \citep{keilholz2017time, preti2016dynamic}. An fMRI measurement
is always only a single realization of the underlying stochastic process and it may therefore show connectivity fluctuations even if the underlying process is stationary \citep{liegeois2017interpreting}. However, obtaining multiple realizations of the
exactly same process is impossible --  the measurements of different subjects as well as the measurements of same subject at different times are different processes \citep{liegeois2017interpreting}. Therefore,
it is hard to construct a proper null model for evaluating the statistical significance of dynamic functional connectivity \citep{liegeois2017interpreting, miller2017resting}. We do not use a stationary null model in the present study, similarly to many other studies. 

If one wants to investigate in detail whether the observed changes in spatial consistency are meaningful, two different paths can be taken. First, testing the results against a statistically rigorous null model would ensure their significance.
Second, the neurophysiological meaning of spatiotemporal consistency could be addressed by comparing consistencies obtained for data measured during rest and during different tasks: obtaining similar changes in response to a common
stimulus in a group of subjects can be considered as indicative of some real underlying mechanism, even if no formal null model is applied.

\subsection{Functionally homogeneous and inhomogeneous ROIs have both turnover in their network neighborhoods} 

The structure of functional brain networks changes with cognitive tasks \citep{braun2015dynamic, gottlich2017viewing, chan2017resting}, and increased local connectivity can be associated with
increased activity and cognitive demand \citep{zang2004, jiang2016regional, hearne2017reconfiguration}. 
If the temporal variation in spatial consistency is
related to changes in brain function, one would expect to see simultaneous changes in network structure as well. Indeed, there was clear turnover in the closest neighborhoods of ROIs, "closest" being defined as the 35 most 
strongly connected neighbors. This turnover was lower for ROIs with high spatial and spatiotemporal consistency; however, even these ROIs changed up to half of their closest neighbors between consecutive time windows.
This indicates
that the local structure of functional brain networks truly changes on short time scales. Further, the network turnover investigated here only quantifies the changes in the identities of the closest neighbors but does not take
into account changes in connection strengths within the closest neighborhood. Therefore, prominent changes may take place in the ranks of the closest neighbors of even a ROI with moderately low network turnover.

We saw that network turnover varies across ROIs. One may speculate about how this variation may relate to the ROIs' different functional roles. It is possible that some ROIs need a diverse and varying set of connections for performing 
their cognitive tasks, while
others require a stable set of neighbors. However, there may be a more straightforward explanation for the variation in network turnover. 
The ROIs with the highest network turnover are subcortical and cerebellar areas that also have low spatiotemporal and spatial consistency. The SNR of the signals originating from these ROIs
is known to be low in fMRI measurements \citep{glasser2016multi}. This may partially explain their low spatiotemporal and spatial consistency and also suggests that their connectivity may be inaccurately mapped. Therefore, 
their extreme network turnover may be partially explained by measurement noise.

We obtained temporal changes in both the spatial consistency and the closest neighborhoods of ROIs. However, we did not investigate the exact timing of these changes. If neighborhood turnover and variation in spatial consistency 
are both 
caused by changes in brain activity, these changes should be more or less simultaneous. This would result in a temporal correlation between the variation of spatial consistency and the neighborhood turnover. The datasets used in the 
present study -- free music listening
and resting state -- may not necessarily be optimal for this kind of investigation. Cognitive responses to the music may differ 
between subjects, and in the resting state subjects are instructed to let their
mind wander uncontrolled. A more detailed investigation of the connection between spatial consistency and turnover would require a dataset with more control on the timing
of putative activity changes. This could be achieved with the traditional block design, where stimuli are repeated at set intervals and the timeline is divided into blocks (see, \textit{e.g.},\citet{tie2009comparison}).
However, the shortness of the blocks is problematic from the viewpoint of network studies: reliable estimation of a functional brain network
requires time series significantly longer than typical block lengths.

\subsection{The internal structure of a ROI may relate to its functional role}

Functional networks are constructed using only the averaged ROI time series, and the only feature that is used in any subsequent analysis is the ROI's location on the brain surface. At the same time, their size, shape, and in particular internal
connectivity are typically ignored. This view of ROIs as featureless entities may, however, be largely oversimplified. 
We found rich, time-dependent structure of voxel-level correlations inside ROIs. Considering the complexity of the brain and the small number of ROIs and their connections to which this complexity is reduced, this is not surprising at all.

The ROIs that we investigated have very different-looking internal structures. These are not necessarily reflected in their consistency measures; in Fig.~\ref{fig:results-corrmats} the two uppermost ROIs have high spatiotemporal consistency, but their correlation matrices display different kinds of structures. The same applies to the two low-consistency bottom rows. 

Why do ROIs have different kinds of internal structures? A plausible hypothesis is that correlation structure inside a ROI relates to ROI's functional role. Let us consider local and connector hubs 
\citep{guimera2005functional,bullmore2009complex} as an example. Local hub nodes are central in their local network modules and have only few connections to nodes outside of their own module, whereas connector hubs act as bridges between different
modules. So, could one separate local and connector hubs from each other in terms of their internal structure? Local hubs are connected only to a relatively stable neighborhood; one might expect that the voxel-level correlation 
distribution inside them is relatively uniform, and periods of high and low voxel-level correlations reflect changes in the activity of the ROI. Connector hubs, on the other hand, need to be able to connect to several different network modules;
an internal structure of diverse subareas could help in this. 

\subsection{Can  brain networks be modelled by static nodes?}

When ROIs are used as nodes of functional brain networks, the brain is assumed to contain a set of static functional areas. An optimal parcellation of the brain then maps to these areas, resulting in functionally 
homogeneous ROIs. If the static-area assumption holds, low functional homogeneity of ROIs then only tells about inaccuracies in ROI definitions which can be corrected by a more accurate parcellation scheme.

Time-dependent changes have been reported in the module structure of functional brain networks at the ROI level \citep{khambhati2017}. Our results suggest that similar changes also occur  in the voxel-level correlation 
structure inside ROIs.
The dynamicaly changing internal connectivity of ROIs appears to challenge the assumption of static functional areas. Because the ROIs of multiple different parcellations have time-varying structure, it seems plausible that
the changing internal connectivity is not a technical issue that can be fixed by an elaborate parcellation scheme. Instead, it may be a genuine feature and related to how the brain works. If so, it may even be impossible to define ROIs in a way that would make them persistently homogeneous.

Many problems caused by the functional inhomogeneity of ROIs can be overcome by using fMRI measurement voxels as nodes of brain networks \citep{hayasaka2010comparison, fornito2013graph}. However, there is evidence for existence of 
functional areas larger than single voxels \citep{wig2011concepts, shen2013}, which motivates investigating brain networks above the level of voxels as well. For example \citet{preti2017dynamics} have suggested an approach
for parcellating the brain based on the dynamic connectivity of voxels; these parcels would be an interesting option for defining network nodes. Similarly, the negative correlation we observed between network turnover and
spatial consistency could be used for defining ROIs: defining ROIs as clusters of voxels that have minimal network turnover should produce ROIs with reasonable spatial consistency.

An optimal network model of the brain should measure
the dynamic connectivity between clusters of voxels and also quantify the changing internal structure of these clusters. In the coarse-graining approach by \citet{kujala2016graph}, self-links are used to model changes in internal
connectivity. As long as static sets of nodes are used to model the time-dependent connectivity of the human brain, outcomes of brain network analysis may be surprisingly inaccurate. Therefore, network neuroscience would greatly
benefit from node definition strategies tailored for dynamic networks.

\section{Materials and Methods} \label{methods}

\subsection{Subjects}

fMRI data of 13 subjects (7 female, 6 male, age 28.70$\pm$10.17 years, mean$\pm$SD, 1 left-handed, 12 right-handed) were used in this present study. The data were collected as a part of a study of functional connectivity during 
music 
listening, containing both musicians and non-musicians freely listening to music, and have been earlier described in \citet{alluri2015musical, burunat2015action, alluri2017connectivity}. For the present study, we used the parts of the 
dataset that were
readily available at the Department of Neuroscience and Biomedical Engineering, Aalto University. The subjects used in the present study 
were considered as non-musicians, \textit{i.e.} had no formal musical training.

All participants signed an informed consent on arrival to the laboratory and received compensation for the use of their time. All experimental procedures for this study, included in the broad research protocol termed “Tunteet”, 
were approved by the Coordinating Ethics Committee of the Hospital District of Helsinki and Uusimaa (the approval number 315/13/03/00/11, obtained on March the 11th, 2012). All procedures were conducted in agreement with
the ethical principles of Declaration of Helsinki.

\subsection{Data acquisition} \label{methods:acquisition}

Functional magnetic resonance imaging (fMRI) data were acquired using a 3T MAGNETOM Skyra scanner (Siemens Healthcare, Erlangen, Germany) and a standard 32-channel head-neck coil in the AMI Centre (Aalto Neuroimaging, Aalto University,
Espoo, Finland). A T2*-weighted whole-brain EPI sequence was measured with the following parameters: TR = 2s, 33 oblique slices, TE = 32ms, flip angle = 75$^{\circ}$, voxel size = 3$\times$3$\times$4mm$^{3}$, 
FOV = 192 $\times$ 192mm$^{2}$, matrix size = 64 $\times$ 64. T1-weighted structural magnetic resonance images (MRI) were acquired with the following parameters: 176 slices, FOV = 256$\times$256mm$^{2}$, matrix size = 256 $\times$ 256, slice
thickness = 1mm.

During the measurement, subjects were instructed to fix their gaze on the screen and actively listen to a musical stimulus (Adios, Nonino by Astor Piazzolla) via MR-compatible insert earphones. Foam was used to suppress the noise caused by the 
imaging gradients. Duration of the stimulus, and therefore of the measured time series, was 8.13 minutes (244 samples).

\subsection{Preprocessing of the data} \label{methods:preprocessing}

The data were preprocessed with FSL software (\url{www.fmrib.ox.ac.uk}, version 5.0.9) and custom in-house MATLAB code (BRAMILA pipeline v2.0, available at 
\url{https://version.aalto.fi/gitlab/BML/bramila}) following the standard fMRI preprocessing steps. This included EPI slice time correction as well as head motion correction using 
MCFLIRT. The data were coregistered to the Montreal Neurological Institute (MNI) 152 2mm standard template using FLIRT two-step procedure where the EPI volumes were first registered to 
the anatomical image of participant’s brain (9 degrees of freedom) and the participant’s anatomical image was then registered to the standard template 
(12 degrees of freedom). No spatial smoothing was applied, but a 240-sec-long cubic Savitzky-Golay filter \cite{ccukur2013attention} was used to remove scanner drift, and the BOLD 
time series were filtered using a Butterworth bandpass filter at 0.01-0.08 Hz. For increased control of motion and physiological artefacts, 24 motion-related 
regressors, signal from deep white matter, ventricles and cerebrospinal fluid were regressed out of the BOLD time series \cite{power2014methods}. 

Voxels with over 70\% of their variance explained by motion or signal from tissues other than the grey matter were removed from the analysis.

\subsection{Regions of Interest}

After preprocessing, we divided the cortex, subcortical areas, and cerebellum into Regions of Interest (ROIs). We used ROIs from three commonly-used parcellations: the connectivity-based Brainnetome
atlas as well as the anatomical Automated Anatomical Labeling (AAL) and HarvardOxford (HO) atlases. In order to build the group-level mask for each of the parcellations, we used the subject-wise analysis masks
obtained as a part of the preprocessing pipeline to account for individual differences in anatomy, and included in the group-level mask only voxels that were present in the analysis masks of all subjects.
Voxel-wise time series were extracted for each ROI and ROI-wise time series were obtained as an average of these voxel-wise time series within each ROI:

\begin{equation}
 X_I(t)=\frac{1}{N_i}\sum_{i\in I}x_i(t),
\end{equation}
where $X_I(t)$ is the time series of the focal ROI $I$, $N_I$ is its size defined as the number of constituent voxels, $x_i(t)$ is the time series of voxel $i$, and summation is over voxels $i$ in the focal ROI.

Some of the parcellations used in this study, in particular AAL and Craddock 200/400, are known to show rather low mean functional homogeneity across ROIs \citep{gordon2014generation}.
However, the ROIs of these parcellations are commonly used as nodes of functional brain networks, and therefore we have chosen to include them in our study.

\subsubsection{Brainnetome atlas}

The Brainnetome atlas \citep{fan2016human} is based on combination of structural and functional connectivity measured by multimodal imaging techniques. In the present study, we used 246 Brainnetome ROIs.
210 of these ROIs were located in the cerebral cortex, while 36 ROIs covered subcortical gray matter. Note that the Brainnetome atlas does not include cerebellar ROIs.

Size of the Brainnetome ROIs varied between 6 and 1102 with a median of 414. Mean ROI size was 424.02$\pm$222.76 (mean$\pm$SD).

\subsubsection{Craddock 200/400}

The connectivity-based Craddock parcellations \newline
\citep{craddock2012whole} have been obtained by applying a two-level normalized cut spectral clustering algorithm on the voxel-level resting-state connectivity matrix. In the present study, we investigate the Craddock 200 and Craddock 400 parcellations that contain 200 and 392 ROIs, respectively, covering the cerebral cortex, subcortical areas, and cerebellum.

The sizes of the Craddock 200 ROIs varied between 202 and 1239 with a median of 706.5. The mean ROI size was 689.96$\pm$168.12 (mean$\pm$SD). The sizes of the Craddock 400 ROIs varied between 56 and 600 with a median of 354.5. The mean ROI size was 352.02$\pm$88.787 (mean$\pm$SD).

\subsubsection{Automated Anatomical Labeling atlas}

AAL \citep{tzourio2002automated} is an anatomical atlas that has been obtained by parcellating a spatially normalized high-resolution single-subject structural volume based on the main sulci. After the parcellation, each ROI has been automatically
associated with a label. We used 116 AAL ROIs, 90 of which were located in the cerebral cortex, 8 in the subcortical gray matter, and 18 in the cerebellum.

Size of the AAL ROIs varied between 44 and 4370 with a median of 1158.5 and a mean of 1366.01$\pm$929.64. 

\subsubsection{HarvardOxford atlas}

The HO atlas (\url{http://neuro.debian.net/pkgs/fsl-harvard-oxford-atlases.html}, \citet{desikan2006automated}) is a probabilistic atlas, where the brain is divided into ROIs based on macroanatomical boundaries. We used
HO ROIs at the probability level of 30\%, meaning that each voxel belonged to the ROI it is associated with in 30\% or more of the subjects used to construct the atlas. We used 138 HO ROIs, out of which 96 were located in the cerebral cortex,
15 covered subcortical gray matter, and 27 were located in the cerebellum. Note that one of the cerebellar ROIs of the HO atlas (Vermis Crus I) is not defined at the probability level of 30\%. Therefore, this ROI is not included
in the present study.

Size of the HO ROIs varied between 28 and 5578 with a median of 633.5 and a mean of 915.63$\pm$921.83 (mean$\pm$SD).

\subsection{Network extraction} \label{methods:network}

In order to construct the dynamic functional brain networks, the time series were divided into 
time windows of 80 samples. This corresponds to a duration of 160s. The consecutive time windows had an overlap of 50\%. This resulted in a total of five time windows along the duration of 
the scan.

The window length was selected so that we were able to investigate the changes of spatial consistency and local network structure (see below) across as many windows as possible, but the values of spatial consistency
were not affected by the short window length. The window length that we used was selected so that further increasing it did not increase the value of spatial consistency obtained in the window 
(see Supplementary Methods for details on selecting the window length). It has been suggested that the time window length should be equal or larger than $1/f_{min}$ where $f_{min}$ is the lowest
signal frequency present in the data \citep{leonardi2015spurious, shakil2015frequency, shakil2017parametric}; the selected window length fulfills this condition. Further, time windows of similar length have been used for constructing
dynamic brain networks in the literature \citep{bassett2011dynamic, bassett2013task}.

We computed the ROI-level adjacency matrix $A$ separately in each of the time windows. The elements $A_{IJ}$ of the adjacency matrix quantified the connectivity between ROIs $I$ and $J$,
defined as Pearson correlation coefficient between their ROI time series. The diagonal of the adjacency matrix was set to zero in order to remove self-links. No thresholding of the correlation 
values was performed at this stage.

\subsection{Spatial and spatiotemporal consistency} \label{methods:spatialconsistency}

For quantifying the functional homogeneity of the ROIs, we used spatial consistency that we have introduced in \citet{korhonen2017consistency}. The spatial consistency $\phi_{spatial}(I)$ of ROI $I$
is defined as the mean Pearson correlation coefficient between the time series of voxels within the ROI:

\begin{equation}\label{eq:spatial-consistency}
 \phi_{spatial}(I)=\frac{1}{N_I(N_I-1)}\sum_{i,i^{\prime}\in I}C(x_i(t),x_{i^{\prime}}(t)),
\end{equation}
where voxels $i$ and $i^{\prime}$ belong to ROI $I$ and $C$
denotes the Pearson correlation coefficient. 

We calculated spatial consistency of all ROIs separately in each time window. For quantifying the variation of spatial consistency across time, we defined spatiotemporal consistency for ROI $I$ as

\begin{equation}\label{eq:stability-of-spatial}
  \phi_{spatiotemporal} = \frac{N_t(N_t-1)}{2\sum_{t<t^{\prime}}\frac{\left|\phi_{spatial}(I,t)-\phi_{spatial}(I,t^{\prime})\right|}{\phi_{spatial}(I,t)}},
\end{equation}
where $N_t$ is the number of time windows, $\phi_{spatial}(I,t)$ denotes spatial consistency of ROI $I$ in time window $t$, and the summation is over all possible pairs of time windows $t$ and $t^{\prime}$. 
As an alternative measure of stability, we used inverse of standard deviation ($1/SD$) calculated over time windows (see Supplementary Results for details).

\subsection{Network turnover} \label{methods:networkconsistency}

The stability of the local network structure around a node was evaluated by computing turnover of its closest neighborhood \citep{saramaki2014persistence, centellegher2017personality}. In this measure, each node was treated as 
an ego that has a certain set of links to 
other nodes referred to as alters. These alters may change across time. We calculated the Jaccard index of the node's 35 top neighbors between consecutive time windows to quantify the amount of change in the closest neighborhood.
This resulted in four Jaccard index values, one for each pair of consecutive time windows. We then defined the network turnover of node $I$ as
\begin{equation}\label{eq:network-consistency}
 \delta_{network}(I)=1-\mu_I^{Jaccard},
\end{equation}
where $\mu_I^{Jaccard}$ denotes the mean Jaccard index of node $I$ across the time windows.

The behavior of turnover as a function of the size of the neighborhood varies between ROIs, especially in small neighborhoods (for details, see Supplementary Methods). We selected the neighborhood
size so that this variation associated with small neighborhoods has stabilised but the trivial decrease of turnover due to large neighborhood size had not yet started.

\subsection{ABIDE data}

In order to ensure that our results are not explained by any feature of our in-house dataset, we repeated all analyses of the present study for a secondary, independent dataset to which we from now on will refer as the 
ABIDE dataset was part of the Autism Brain Imaging Data Exchange I (ABIDE I) initiative
\citep{di2014autism} and contained resting-state data of 28 healthy adult subjects. Importantly, data of these subjects were collected with the same TR as our in-house data ($TR=2.0s$); differences in TR could have caused unexpected
effects in correlation-based measures. Details about the ABIDE data can be found in Supplementary Methods.

\section*{Acknowledgements}

We acknowledge the computational resources provided by the Aalto Science-IT project. We wish to thank the Danish National Research Foundation (project DNRF 117) for financial support. Moreover, 
we are grateful to Brigitte Bogert, Benjamin Gold, Marina Kliuchko, David Ellison, Taru Numminen-Kontti, 
Mikko Heim\"al\"a, Jyrki M\"akel\"a, Marita Kattelus, and Toni Auranen for their contribution to data collection. We wish also to thank Petri Toiviainen, Vinoo Alluri, and Iballa Burunat for their help in several aspects of this project.

\section*{Supplementary information}

Supplementary information for this article is available at \url{https://github.com/onerva-korhonen/ROI_consistency}. In the Supplementary Methods section, we describe in detail the selection of time window length and closest
neighborhood size as well as the ABIDE dataset. In Supplementary Results, we cover the results obtained by repeating the analysis using the Automated Anatomical Labeling (AAL), HarvardOxford (HO), and Craddock 200/400
atlases as well as using the ABIDE dataset. Further, we investigate an alternative definition of spatiotemporal consistency, the role of the time window length, and the connection between spatial consistency and time-resolved
functional connectivity.

\bibliographystyle{apacite}

\bibliography{consistency_and_activity_bibliography}

\end{document}